\documentclass[twocolumn,preprintnumbers,amsmath,amssymb,floatfix,showpacs,prb]{revtex4}

\usepackage{dcolumn}
\usepackage{bm}
\usepackage{graphicx}

\begin{document}


\title{Ground State Properties of Antiferromagnetic Chains with
  Unrestricted Spin: Integer Spin Chains as Realisations of the $O(3)$
Non-Linear Sigma Model}

\author{F. D. M. Haldane}
\altaffiliation{Present address 
: Department of Physics, University of
  Southern California, University Park, Los Angeles CA 90007
  [\textit{Note: this was correct in 1981, but is no longer the
    author's address}.]}
\affiliation{Institut Laue-Langevin,\\
156X, 38042 Grenoble, France}

\date{July 1981}

\begin{abstract}
A continuum limit treatment of planar spin chains  with arbitrary $S$
is presented.
The difference between integer and half-integer spins is emphasised.
While isotropic half-integer spin chains are gapless, and have
power-law decay of correlations at $T$ = 0 with exponent $\eta$ = 1,
integer spin systems have a singlet ground state with a gap for $S=1$
excitations and exponential decay of correlations.   The easy-plane to
easy-axis transition is described.\\
\\
\noindent
\textit{Note:  this is a verbatim transcription of ILL preprint
  SP-81/95,
 which is cited in a number of places, but was rejected for
 publication in 1981, and remained unpublished
  in this original form, which gives somewhat  different arguments for the
central  result as compared to the later paper finally published in
 Phys. Lett. 93A,  464 (1983);  
many thanks to Jen\H{o} S\'{o}lyom for preserving this historical document.
The author's current address (November 
2016)  is: Department
  of Physics, Princeton University, Princeton NJ 08544-0708, USA.
}
\end{abstract}
\maketitle

One dimensional quantum spin chains are currently the subject of much
study.
In this note, I outline some new results on axially-symmetric spin
chains, without restrictions on $S$, that confirm and extend earlier
results\cite{ref1,ref2} restricted to spin $S$ = $\frac{1}{2}$, and
lead to an unexpected conclusion: while \textit{half-integral spin}
isotropic antiferromagnets have a gapless linear spin wave spectrum
and \textit{power-law decay} of ground state correlations 
$\langle\vec S_n\cdot \vec S_0\rangle $ $\sim$ $(-1)^nn^{-1}$,
\textit{integral spin} system have a singlet ground state, with a gap
    for
``massive'' $S$ =1 elementary excitations, and \textit{exponential
  decay}
of ground state correlations:
$\langle\vec S_n\cdot \vec S_0\rangle $ $\sim$
$(-1)^nn^{-\frac{1}{2}}\exp( -\kappa n)$.   Indeed, as will be
outlined, the isotropic integer spin system can be related to the $O(3)$
coupled rotor chain, or ``non-linear sigma model''\cite{ref3}, and
thus to the 2D classical Heisenberg model.  The role of quantum
fluctuations controlled by $S^{-1}$ is analogous to thermal
fluctuations in the latter model.

I will discuss the $XY$-Heisenberg-Ising system, with all components
of exchange antiferromagnetic: with $\lambda \ge 0$,
\begin{equation}
H = |J|\sum_n\left ( S^x_nS^x_{n+1} + S^y_nS^y_{n+1} + \lambda
  S^z_nS^z_{n+1}\right ).
\label{eq1}
\end{equation}
In the easy-plane limit $(\lambda \ll 1)$, a description in terms of
quantum
\textit{action-angle} variables $N$, $\phi$ (where $[N, \exp (i\phi)]$
$=$
$\exp (i\phi)$) is appropriate: $N_n$ = $S + S^z_n$, $S^+$ = 
$(-1)^n(S+ S^z_n)^{\frac{1}{2}} \exp (i \phi_n)
  (S-S^z_n)^{\frac{1}{2}}$. This representation, similar in spirit to
  the Holstein-Primakoff boson representation appropriate in the easy-axis limit,
  is exact\cite{ref4}. 
The eigenvalues of $N_n$ are integers.  Physical states are restricted
to the subspace $0 \le N_n \le 2S$.

It will prove useful to introduce a dual ``angle'' variable
$\theta_{n+\frac{1}{2}}$ defined on bonds, so $2\pi S_n^z$ =
$\theta_{n+\frac{1}{2}} - \theta_{n-\frac{1}{2}}$.  In the limit of
sufficiently strong easy-plane anisotropy, $\phi_n$ will exhibit
short-range order, and vary slowly along the chain.  I will develop a
continuum description in terms of fields $\phi(x)$ and $\theta(x)$,
where for lattice spacing $a$ and $x_n$ $\equiv$ $na$, $\phi_n \sim
\phi(x_n)$, and $\Pi(x_n)$ = $[\theta(x_{n+\frac{1}{2}}) -
\theta(x_{n-\frac{1}{2}})]/2\pi a  -   S^z/L$ is conjugate to $\phi(x_n)$. 
This description is appropriate when zero-point fluctuations of
$\phi_n$ with respect to its neighbours are small compared to $\pi$.
In this small-fluctuation regime, the locally-periodic character of
$\phi_n$ and the related discretisation of the spectrum of $S^z_n$ are hidden.

Noting that periodic boundary conditions allow the field $\phi(x)$
to increase by $2\pi$ times an integer $J$ around the ring of length
$L$, the fields $\theta(x)$ and $\phi(x)$ can be represented as
\begin{eqnarray}
\theta(x) &=& \theta_0 + 2\pi S^zx/L - i \sum_q \alpha_q^+e^{iqx}
{\rm sgn}(q) (b^{\dagger}_q + b_{-q}), \nonumber \\
\phi(x) &=&\phi_0 + 2\pi Jx/L -i \sum_{q}
  \alpha^-_qe^{iqx}(b^{\dagger}_q - b_{-q}),
\end{eqnarray}
where $\alpha^{\pm}_q$ = $(2L\sin (\frac{1}{2}|q|a)/\pi
a)^{-\frac{1}{2}}\exp (\pm \varphi(q))$, and $\varphi(q)$ is a free
Bogoliubov transformation parameter.   $b^{\dagger}_q$ are boson
creation operators labelled by $q$ = $2\pi n/L$, $q\ne 0$,
$|q| < \pi/a$.   $\phi_0$ is the global spin
angle conjugate to total azimuthal spin  $S^z$; similarly, $\theta_0$
is the angle conjugate to the action-type variable $J$.

The momentum operator $P$ is given by
\begin{equation}
P = (\pi + 2\pi J/L)(S^z + SL/a)/a + \sum_qqb^{\dagger}_qb_q.
\label{eq3}
\end{equation}
The first term provides a global rotation of $\phi_0$ by $\pi + 2\pi
J/L$.   This is required in addition to a shift of the pattern of fluctuations
for a translation of the spin configuration by one lattice spacing.

The representation of $S^z_n$ in a way that reflects its discrete
spectrum is crucial.  The conserved magnetisation density can be
regarded as residing in a fluid of ``magnon'' excitations about the
fully-aligned state $S^z_n$ = $-S$, each magnon carrying $S^z$ = $+1$.
 As $n$ increases, $\theta_{n+\frac{1}{2}} + 2\pi S(n+\frac{1}{2}) $
 increases
monotonically by $2\pi$ each time such a magnon is passed.  In the
continuum field description, the magnons will be taken to be located
at the points where $\theta(x) + 2\pi Sx/a$ is an integer multiple of
$2\pi$.  The magnon density operator is thus a sum of unit-weight
delta-functions: $\rho(x)$ =
\begin{equation*}
\sum_n \delta[\theta(x) + 2\pi Sx/a -2n\pi] [\nabla\theta(x) + 2\pi
S/a].
\end{equation*}
This can be written as $\rho(x)$ = $S/a$ + $(2\pi)^{-1}\nabla \tilde
\theta(x)$,
\begin{equation}
\tilde \theta (x) = \theta(x) + \sum_{m\ge 1} 2\sin[m\theta(x) + 2\pi
m Sx/a)]/m.
\label{eq4}
\end{equation}
The spin operator $S^z_n$ is then given by the integrated magnon
density in the unit cell $x_{n-\frac{1}{2}} \le x \le
  x_{n+\frac{1}{2}}$; this immediately gives $\theta_{n+\frac{1}{2}}$
  = $\tilde \theta(x_{n+\frac{1}{2}})$.   $S^z_n$  given in terms of
  the operators $\theta_{n+\frac{1}{2}}$ constructed out of the
  continuum fields $\theta(x)$ has the required discrete spectrum.
For calculation of long-distance correlations, a gradient expansion
in $\theta(x)$ can be used to obtain an approximate \textit{local}
form for $S^z_n$.  This form depends on whether $S$ is integral or 
half-integral through the term $2\pi mSx/a$ in (\ref{eq4}).
For \textit{integral} $S$,
\begin{equation}
S^z_n \sim a [S^z/L + \Pi(x_n)]\{1 + \sum_{m \ge 1} 2\cos
[m\theta(x_n)]\};
\label{eq5}
\end{equation}
for \textit{half-integral} S, an oscillatory term is present: $S^z_n$
$\sim$ $S^z_{n1} + (-1)^n S^z_{n2}$, where
\begin{eqnarray}
S^z_{n1} &=& a[S^z/L + \Pi(x_n)]\{1 + \sum_{m\ge 1}
             2\cos[2m\theta(x_n)]\}, \nonumber \\
S^z_{n2} &=& \frac{1}{\pi}\sum_{m\ge 0} 2\sin[(2m+1)\theta(x_n)]/(2m+1).
\label{eq6}
\end{eqnarray}

If, following Villain\cite{ref4}, the long-wavelength approximation
$S^z_n$ $\sim$ $a[S^z/L + \Pi(x_n)]$ is made, linearisation of (\ref{eq1})
in $S^z_n$ and $(\phi_{n+1} - \phi_n)$ when $\lambda < 1$ leads to the
effective Hamiltonian
\begin{equation}
H = {\textstyle\sum_q} \omega(q) b^{\dagger}_qb_q + (\pi v_s/L)[\eta(S^z)^2 +
\bar{\eta} J^2],
\label{eq7}
\end{equation}
where $\eta$ = $\bar{\eta}^{-1}$ = $\exp[-2\varphi(0)]$,
$\exp [2\varphi(q)]$ = $2\pi S[2 + 2\lambda\cos(qa)]^{\frac{1}{2}}$.
The spin-wave spectrum is linear as $q\rightarrow 0$:  $\omega(q)$
$\sim$ $v_s|q|$; $\omega(q)$ =
$2JS\sin(\frac{1}{2}|q|a)[2 + 2\lambda \cos(qa)]^{-\frac{1}{2}}$ - note
the
\textit{soft mode} at the Brillouin zone edge $q$ = $\pi/a$ $\equiv$
$\frac{1}{2}G$.  The linearisation is valid for large $S$, when
non-linear zero-point fluctuations are suppressed; however, for general
$S$, provided such fluctuations do not lead to breakdown of the
short-range order of $\phi_n$, a renormalisation procedure should
yield an effective Hamiltonian of form (\ref{eq7}), but with
renormalised parameters $\varphi(q)$, $\omega(q)$.   This viewpoint
has been advanced in Ref.(\onlinecite{ref5}), and is supported by studies of
exactly-soluble models. The state described by (\ref{eq7}) may be
called a ``spin fluid'' state.   The magnon current $j$ =
$(2\pi)^{-1}(d\theta_0/dt)$ = $\bar{\eta}v_s(J/L)$ is conserved at low
energies.

Correlation functions are easily evaluated in the fluid state.  The
momentum associated with the current excitation $\Delta J$ = 1 is
$SG + 2\pi S^z/L$.   When $S^z$ $\ne$ $ 0$, harmonics of this
wavevector
appear in the correlations through sine and cosine terms in
(\ref{eq5}) and (\ref{eq6}).   However, I will describe only the case
$S^z$ = 0.

For {\textit{integer} spins, the current excitation carries momentum
0 (${\rm mod.} \,G$).
However, oscillatory terms
with exponential decay constant $\kappa_0$ (where $\omega(\frac{1}{2}G
+ i\kappa_0)$ = 0) are still present, due to the soft-mode in the
spin-density
fluctuation spectrum at the Brillouin zone edge; these arise from the
branch cut in $\exp(2\varphi(q))$ at complex $q$.
In the linearised approximation, $\cosh (\kappa_0 )$ = $1/\lambda$.
For $n \gg 1$, $\langle S^z_nS^z_0\rangle$
$\sim$ $A(-1)^nn^{-\frac{1}{2}}\exp (-\kappa_0n) + \bar{\eta}
(2\pi n)^{-2}(1 + B n^{-\bar{\eta}})$,
$\langle S^+_nS^-_0\rangle $ $\sim$ $C(-1)^nn^{-\eta} +
Dn^{-\frac{3}{2}}
\exp(-\kappa_0 n)$, where $A$ - $D$ are constants depending on
short-wavelength structure.

For \textit{half-integral} spins, the current excitation carries
momentum $\frac{1}{2}G$, and controls oscillatory behaviour, masking
the soft mode.
For $n \gg 1$, $\langle S^z_nS^z_0\rangle$ $\sim$
$A(-1)^nn^{-\bar{\eta}}$ + $\bar{\eta} (2\pi n)^{-2} (1 + B
n^{-4\bar{\eta}})$, $\langle S^+_nS^-_0\rangle$ $\sim$
$Cn^{-\eta}[(-1)^n + D n^{-\bar{\eta}}]$.   These expressions agree
with the $S$ = $\frac{1}{2}$ results previously obtained with the various
fermion
representations\cite{ref1,ref2}.

If the full forms (\ref{eq5}), (\ref{eq6}) for $S^z_n$ are used, terms
  involving $\cos (m\theta(x_n))$ are seen to be present in the
  Hamiltonian.  These terms can be regarded as Umklapp terms, as they
  allow destruction of current quanta.  When $S^z$ = 0, individual
  current quanta can be destroyed in integral spin systems (by a
  $2\pi$ rotation of a spin), while in  half-integral spin systems,
  they can only be destroyed in \textit{pairs} (by a local $4\pi$
  rotation).   The long-wavelength fluctuation part of the
  Hamiltonian has the form (\ref{eq8}), with $c$ = $v_s/4\pi$:
\begin{equation}
H = c\int dx\, [\bar {\eta}(\nabla \phi)^2 + \eta(\nabla \theta)^2 +
\sum_m \gamma_m \cos(m\theta)].
\label{eq8}
\end{equation}
Since $(2\pi)^{-1}\nabla \phi$ is conjugate to $\theta$, this is
easily recognised to be of sine-Gordon type\cite{ref6}, with coupling
parameters
$\beta_m^2$ = $2\pi m^2\bar{\eta}$.
The fluid state is only stable if $\beta_m^2 > 8\pi$, \textit{i.e.},
$\eta$ $<$ $\frac{1}{4}m^2$.
In the fluid state, fluctuations of $\nabla \phi(x)$
are small compared to the conjugate fluctuations of $\theta(x)$.
As $\lambda$ increases, $\eta$ and the fluctuations of $\nabla \phi(x)$
increase, while those of $\theta(x)$ decrease.   Eventually, the
fluctuations
of $\theta(x)$ are too small to prevent pinning by the cosine
potential, and those of $\nabla \phi(x)$ are sufficiently large that
the
local periodicity of $\phi_n$ is restored.

For \textit{integer spins}, the $m=1$ process is present.  $\eta$
reaches its limiting value of $\frac{1}{4}$ at some critical value
$\lambda_{c1} < 1$, at which the correlations still have easy-plane
character.
For $\lambda > \lambda_{c1}$, $\theta(x_n)$ is pinned to values 0
(mod.$\,2\pi$), and $2\pi$ fluctuations of $\phi_n$ are important.  The
resulting state may be described as a pinned spin-density wave with
the \textit{same} periodicity as the lattice, so no broken symmetry
is present.    
Its excitations are ``topological solitons'' where
$\theta(x)$ slips by $2\pi$ (one magnon), carrying $S^z$ = $\pm 1$,
and intrinsic momentum $\frac{1}{2}G$  (from (\ref{eq3})).
Predicted\cite{ref7} correlations in the \textit{singlet ground state}
phase with $\lambda > \lambda_{c1}$ decay as 
$\langle S^z_nS^z_0\rangle$ 
$\sim$
$A(-1)^nn^{-\frac{1}{2}}\exp(-\kappa_0n)$
+ $n^{-2} [B\exp (-2\kappa_0n] + C\exp (-2\kappa_1n)]$,
$\langle S^+_nS^-_0\rangle$ $\sim$ $D(-1)^nn^{-\frac{1}{2}}\exp
(-\kappa_1n)$
+ $En^{-2}\exp(-(\kappa_0 +\kappa_1)n),$ where $\kappa_1$ is the decay
constant
associated with the $S^Z$ = $\pm 1$ soliton dispersion
$\varepsilon_1( q)$: $\epsilon_1(\frac{1}{2}G+ i\kappa_1)$ = 0. 

As $\lambda$ increases, the soliton gap increases, while that of the
soft mode decreases.  At the isotropic point $\lambda$ = 1, $\kappa_0$
= $\kappa_1$, and the soft mode excitation forms a triplet with the
$S^z$ = $\pm 1$ soliton  excitation.   The soft mode excitation has
the lower energy for  $\lambda > 1$, and its gap vanishes at a second
critical point $\lambda_{c2} > 1$, signalling the instability against
the doubly-degenerate Ising-N\'{e}el state.   Though detailed
justification cannot be given here, I note the critical behaviour can
be identified with that of the singlet-doublet transition in the
``$(\phi)^4$'' field theory or 2-D Ising model, just as critical
behaviour at $\lambda_{c1}$ is related to that of the 2-D XY model.
$\kappa_0$ vanishes when $\lambda$ = $\lambda_{c2}$, and 
predicted\cite{ref7} correlations decay as $\langle S^z_nS^z_0\rangle$
$\sim$ $A(-1)^n n^{-\frac{1}{4}}$ + $Bn^{-\frac{3}{2}}$,
$\langle S^+_nS^-_0\rangle$
$\sim$ $n^{-\frac{1}{2}}\exp(-\kappa_1n)(C(-1)^n +
Dn^{-\frac{5}{4}})$.
In the doublet N\'{e}el state when $\lambda$ $>$ $\lambda_{c2}$, the
$S^z$ =0 excitations again develop a gap, and can now  be identified  
as \textit{pairs} of solitons of the N\'{e}el state (which correspond
to configurations like $[+-+0-+-]$, and carry \textit{no}
magnetisation in the integral-spin case).   Predicted\cite{ref7}
correlations now decay as 
$\langle S^z_nS^z_0\rangle$ $\sim$ $(-1)^n\{A + n^{-2}[B
\exp(-2\kappa_0n) +
C \exp(-2\kappa_1n)]\}$ + $ n^{-1}[D \exp (-2\kappa_0n) +
E\exp(-2\kappa_1n)]$,
$\langle S^+_nS^-_0\rangle$ $\sim$
$n^{-\frac{1}{2}}\exp(-\kappa_1n)  [(-1)^nF + Gn^{-1}]$.
As $\lambda$ increases above $\lambda_{c2}$ the excitation energy and
associated decay constant $\kappa_0$ of  the $S^z$ = 0 N\'{e}el
solitons rapidly become larger than the corresponding quantities for
the $S^z$ = $\pm 1$ excitations, now identified as the N\'{e}el
magnons.

The transition from easy-plane 
to easy-axis system is simpler in the case of \textit{half-integer}
spin.  The $m$ =1 Umklapp process is absent, and the $m$ = 2 process
drives an instability 
against a doubly-degenerate N\'{e}el-Ising density-wave state, where
$\theta(x_n)$ alternates between 0 and $\pi$.
The solitons (configurations like $[+-+\sigma-+-]$, $\sigma$  =
$\pm \frac{1}{2}$)
are created in pairs, and carry 
magnetisation $S^z$ = $\pm \frac{1}{2}$.
The critical value of $\eta$ at breakdown of the fluid state is $\eta$
= 1, when the correlations are isotropic, hence the transition occurs
at $\lambda$ = 1.   For $\lambda$ $\gtrsim$ 1, \textit{pairs} of $S^z$
= $\pm \frac{1}{2}$ solitons are the lowest energy excitations, and a
strong enough axial field will drive a second-order transition into a
weakly incommensurate density wave state, which can be considered as a
dilute fluid of $S^z$ = $+\frac{1}{2}$ solitons.  The origin of the
$S^z$ = $\pm 1$ N\'{e}el magnon excitations, which have much lower
energy than the solitons at larger easy-axis anisotropies, is not yet
clarified.  In the strong anisotropy limit, the response to a magnetic
field is a first-order ``spin-flop'' transition.  It seems likely that
the easy-axis N\'{e}el magnon is connected with the easy-plane
zone-edge soft
mode; it is not clear whether these excitations have a gap at $\lambda$
= 1, when the soliton gap vanishes.

The soluble $S$ = $\frac{1}{2}$ chain exhibits the $m$ = 2 Umklapp
instability, which was overlooked in the earlier ``fermion
representation'' treatment\cite{ref1}, but first pointed out in
Ref.(\onlinecite{ref5}).  The details of the solution are in precise accord
with the scaling theory of (\ref{eq8})\cite{ref8}.
A feature special to $S$ = $\frac{1}{2}$ is the absence of the
N\'{e}el magnon for $\lambda > 1$, \textit{and} the soft mode for
$\lambda < 1$.    This can be attributed to the hard-core nature of
the quantum of magnetisation.  In the easy-axis case, a treatment in
terms of fermion ``disorder variables'' (solitons) is more appropriate
than a treatment using Holstein-Primakoff boson variables\cite{ref9}.

Finally, I note the similarity between the behaviour described here
for the integral-spin antiferromagnetic chain and that of the $O(3)$
rotor chain, or ``non-linear sigma model'', which can be related to
the classical 2-D $n$=2 vector spin model\cite{ref3}.
The rotor chain Hamiltonian is
\begin{equation}
H = \sum_n {\textstyle\frac{1}{2}}gL_n^2 + 
(\Omega_n^x\Omega_{n+1}^x  +
\Omega_n^y\Omega_{n+1}^y 
+ \lambda \Omega_n^z\Omega_{n+1}^z )
\label{eq9}
\end{equation}
where $\vec L_n\cdot \vec \Omega_n$ = 0,
$\Omega_n^2$ = 1, and the rotor commutation relations are
$\vec L_n \times \vec L_n$ = $i\vec L_n$,
$\frac{1}{2}( \vec L_n \times \vec \Omega_n + \vec\Omega_n
\times \vec L_n) $ = $i\vec \Omega_n$, $\vec \Omega_n \times \vec \Omega_n$ = 0.
In the isotropic case $\lambda$ = 1, this model has a singlet ground
state with a gap for $L=1$ elementary
excitations for all non-zero $g$, and models the 2-D classical
Heisenberg model with $g$ playing the role of temperature\cite{ref3}.
Consider now a spin chain with alternating ferromagnetic and
antiferromagnetic exchange, $H$ = $J\sum_n (-1)^n\vec S_n\cdot
\vec S_{n+1}$.   This model has two spins per unit cell, and to study
it, it is useful to decimate it by blocking the spins into pairs.
If it is chosen to combine ferromagnetically coupled spins, it is
easily seen that the resulting effective Hamiltonian is that of a spin
$2S$ (\textit{i.e.}, \textit{integral} spin) antiferromagnet.
If antiferromagnetically coupled spins are combined, one can introduce
new variables $\vec L$ = $\vec S_1 + \vec S_2$, $\vec \Omega$ =
$(2S)^{-\frac{1}{2}}
(\vec S_1 -\vec S_2)$.  For large $S$ these obey rotor-like
commutation
relations, and an effective rotor-chain Hamiltonian is obtained. The
identification with (\ref{eq9}) provides the connection of the
$\lambda_{c1}$ and $\lambda_{c2}$ instabilities to the 2-D XY and Ising
critical
behaviour, and justifies the discussion of the $\lambda_{c2}$
instability
given above.

\end{document}